\documentclass{PoS}

\usepackage{dsfont}

\newcommand{\beq}{\begin{equation}}
\newcommand{\eeq}{\end{equation}}
\newcommand{\bea}{\begin{eqnarray}}
\newcommand{\eea}{\end{eqnarray}}

\newcommand\eqn[1]{(\ref{#1})}      % parentheses around the LaTex "ref" macro
\newcommand\Eqn[1]{Eq.~(\ref{#1})}  % includes ``Eq.'' in front
\newcommand{\tr}{\hbox{tr}}
\newcommand{\bartheta}{\,\bar{\!\theta}}
\newcommand{\thetab}{{\bartheta}}
\newcommand{\cb}{{\bar c}}

\title{Lifting the Gribov ambiguity in Yang-Mills theories}

\ShortTitle{Lifting the Gribov ambiguity}

\author{\speaker{Julien Serreau}\\
\\
        APC, AstroParticule et Cosmologie, Universit\'e Paris Diderot, CNRS/IN2P3, CEA/Irfu, Observatoire de Paris, Sorbonne Paris Cit\'e,  10, rue Alice Domon et L\'eonie Duquet, 75205 Paris Cedex 13, France\\
        E-mail: \email{serreau@apc.univ-paris7.fr}}

\abstract{We report on the work presented in Ref. \cite{Serreau:2012cg}, where a new one-parameter family of Landau gauges has been proposed for Yang-Mills theories, inspired by an analogy with disordered systems in condensed matter physics. This is based on a particular average over Gribov copies which avoids the Neuberger zero problem of the standard Fadeev-Popov construction. The proposed gauge fixing can be formulated as a local renormalizable field theory in four dimensions and is well-suited for analytical calculations. A remarkable feature is that, for what concerns the calculation of ghost and gauge field correlators, the gauged-fixed action is perturbatively equivalent to a simple massive extension of the Faddeev-Popov action. The renormalization group flow of the theory admits infrared safe trajectories, with no Landau pole. The one-loop calculations of Yang-Mills two-point correlators show remarkable agreement with lattice simulations all the way to the deep infrared.}

\FullConference{Xth Quark Confinement and the Hadron Spectrum,\\
		October 8-12, 2012\\
		TUM Campus Garching, Munich, Germany}

\begin{document}

\section{Introduction}

A proper formulation of gauge fixing in non-abelian gauge theories is a longstanding issue. The existence of Gribov copies for the most common choices of gauges render the standard Fadeev-Popov (FP) construction problematic at least away from the perturbative regime \cite{Gribov77}. For instance, in these gauges the FP construction for a $SU\!(N)$ Yang-Mills (YM) theory discretized on a finite lattice is plagued by the so-called Neuberger zero problem, inherent to any BRST invariant gauge fixing \cite{Neuberger:1986vv}. In the following, we consider the Euclidean theory in the Landau gauge $\partial_\mu A_\mu=0$, which has been the most studied on the lattice. In that case, Gribov copies corresponds to the extrema of the functional ${\cal F}[A]=\int_x\tr \{A^2\}$ along the gauge orbit $A^U$ of a given field configuration $A$.

A way to cope with the Gribov issue is to pick up only one copy for each field configuration. For instance, finding a minimum of ${\cal F}[A^U]$ is a relatively easy task in lattice simulations \cite{Boucaud:2011ug}. However, it is not known how to implement this procedure for continuum approaches, which renders comparisons somewhat tricky \cite{Maashere}. 
Gribov and Zwanziger (GZ) \cite{Gribov77,Zwanziger89} have proposed to restrict the path integral over gauge fields to the first Gribov region, where the FP operator has only positive eigenvalues. This can be approximately formulated in terms of a local, renormalizable field theory \cite{Vandersickel:2012tz} at the price of introducing a collection of auxiliary fields and a dimensionful parameter which controls the (soft) breaking of BRST symmetry. However, the first Gribov region is not exempt of Gribov copies: ${\cal F}[A^U]$ has in fact many minima. A refined version of this proposal, where one introduces phenomenological vacuum condensates, agrees well with lattice data~\cite{Dudal08}.

In \cite{Serreau:2012cg}, we have proposed a different approach to the Gribov problem based on averaging over Gribov copies, instead of trying to single out a particular one, in a way that can be formulated in terms of a local action. The averaging weight can be chosen so as to break the nilpotent BRST symmetry soflty, leading to a perturbatively renormalizable gauged-fixed theory in four dimensions. Furthermore, a non-flat weight lifts the degeneracy between copies and avoids the Neuberger zero problem. The proposal of \cite{Serreau:2012cg} is very much inspired from an analogy between the Gribov problem and that of dealing with  potentials presenting a landscape with (exponentially) large number of nearly degenerate minima in the physics of disordered systems \cite{Tissier:2011zz}. 

In practice, we use the functional ${\cal F}[A^U]$ to weight the various copies. This particular choice allows for an elegant and powerful formulation in terms of a collection of replicated supersymmetric gauged non-linear sigma (NL$\sigma$) models. A remarkable consequence of the symmetries of the theory is that the NL$\sigma$ model sector of the theory essentially decouples in perturbative calculations of YM ghost and gluon correlators. In that case, our gauge-fixing procedure is perturbatively equivalent to a simple massive extension of the FP action in the Landau gauge, which is a particular limit of the Curci-Ferrari (CF) model \cite{Curci76}. This provides a fundamental link between this phenomenological model and YM theories. Recent one-loop calculations of two-point correlators in the CF model have been shown to agree remarkably well with lattice data all the way to the deep infrared \cite{Tissier_10}. Moreover, the presence of a new running parameter (the gluon mass) opens the possibility for infrared safe renormalization schemes, with no Landau pole, as discussed in \cite{Tissier_10}.
In this contribution, I briefly review the main aspects of the proposal of \cite{Serreau:2012cg}, recalling some one-loop results of \cite{Tissier_10}. I discuss some open issues and mention ongoing research in this context.

\section{Weighting Gribov copies}

We consider a $SU\!(N)$ YM theory on $\mathbb{R}^d$, with classical action
\beq S_{\rm
  YM}[A]=\frac{1}{2}\int_x\tr \left\{F^2\right\}\,, 
\eeq 
where $F_{\mu\nu}=\partial_\mu A_\nu-\partial_\nu
A_\mu -ig_0[A_\mu,A_\nu]$ and  $\int_x\equiv\int d^dx$. 
For any operator $\mathcal{O}[A]$, we define the average over the Gribov copies $A^{U_{(i)}}$ of a given field configuration $A$ as:
\begin{equation}
  \label{eq_average_G}
  \langle\mathcal O[A]\rangle=\frac{\sum_i \mathcal O [A^{U_{(i)}}]s(i)e^{-S_{\rm \!W}[A^{U_{(i)}}]}}{\sum_i s(i)e^{-S_{\rm \!W}[A^{U_{(i)}}]}}\,,
\end{equation}
where $s(i)$ is the sign of the functional determinant of the FP operator $-\partial_\mu D_\mu$ evaluated at $A=A^{U_{(i)}}$ and  $S_{\rm \!W}[A]$ is a given positive definite weight functional. Here, the sum runs over all Gribov copies of the Landau gauge, that is over all extrema $U_{(i)}$ of ${\cal F}[A^U]$ for a given $A$. \Eqn{eq_average_G} defines a good gauge fixing in the sense that it does not affect gauge-invariant operators: $ \langle\mathcal O_{\rm inv}[A]\rangle=\mathcal O_{\rm inv} [A]$ for $\mathcal{O}_{\rm inv}[A^U]={\cal O}_{\rm inv}[A]$.
The denominator in \eqn{eq_average_G} is crucial for this property to hold. In the case of a flat weight, $S_{\rm \!W}[A^{U_{(i)}}]={\rm const}$, all copies are degenerate --as in the case of the FP construction-- and this denominator --as well as the numerator in the case of gauge-invariant operators-- vanishes: $\sum_is(i)=0$. For a non-flat weight the degeneracy is lifted and there is no Neuberger zero problem. 

In \cite{Serreau:2012cg}, we choose to weight copies with the functional ${\cal F}[A]$ itself:
\begin{equation}
  \label{eq_z}
S_{\rm \!W}[A]=  \beta_0\int_x\,\tr \left\{A^2\right\}.
\end{equation} 
with $\beta_0>0$. This interpolates between the usual FP construction for $\beta_0\to0$ (see below) and the absolute Landau gauge \cite{Dell'Antonio:1991xt}, which corresponds to selecting the absolute minimum of ${\cal F}[A^U]$, in the limit $\beta_0\to \infty$. We also note that, assuming a gap between the minima of ${\cal F}[A^U]$ and the first saddles, the weight \eqn{eq_z} suppresses copies outside the first Gribov region for not too small $\beta_0$.

Once the average \eqn{eq_average_G} over Gribov copies has been performed for each individual field configuration one performs the usual average over YM field configurations, hereafter denoted by an overall bar. The average of a given operator ${\cal O}[A]$ in our gauge-fixing procedure is thus obtained as a two-step average\footnote{A somewhat similar gauge-fixing has been proposed in \cite{Parrinello:1990pm} where, however, the average was not restricted to Gribov copies in the Landau gauge. This difference is essential, e.g., in making the present proposal renormalizable.}:
\begin{equation}
  \label{eq_av_A}
\overline{\langle\mathcal O[A]\rangle}=\frac{\int\mathcal
  DA\, \langle\mathcal O[A]\rangle \,e^{-S_{\rm YM}[A]}}{\int\mathcal
  DA\, e^{-S_{\rm YM}[A]}}\,.
\end{equation}

\subsection{Field theoretical formulation}

The discrete sums over Gribov copies in \eqn{eq_average_G} can be written as constrained functional integrals over local elements $U(x)$ of the gauge group. The constraint $\partial_\mu A_\mu^U=0$ can be exponentiated by means of a Nakanishi-Lautrup field $h$. Similarly, the corresponding Jacobian multiplied by the sign $s(i)$ in \eqn{eq_average_G} is nothing but the determinant of the FP operator, which can be exponentiated by means of standard FP ghost fields $c,\cb$. Denoting collectively ${\cal V}\equiv(U,c,\cb,h)$, we have:
\beq
\label{eq:fact1}
 \langle {\cal O}[A]\rangle=\frac{\int{\cal D}{\cal V}{\cal O}[A^U]\,e^{-S_{\rm \!GF}[A,{\cal V}]}}{\int{\cal D}{\cal V}\,e^{-S_{\rm \!GF}[A,{\cal V}]}}\,,
\eeq
with the gauge fixing action
\beq
\label{eq:gf}
 S_{\rm \!GF}[A,{\cal V}]=\int_x\tr \Big\{{\beta_0}A^2+2\partial_\mu\bar cD_\mu c+2ih\partial_\mu A_\mu\Big\}_{A=A^U}.
\eeq
The latter presents a number of linear and non linear symmetries, including a generalized non-nilpotent BRST symmetry. These are most easily seen by introducing the matrix superfield 
\begin{equation}
  \label{eq_susy}
  \mathcal V(x,\theta,\bartheta)=e^{ig_0\left(\bartheta c+\bar
    c\theta+\bartheta\theta \tilde h\right)}U
\end{equation}
living on a superspace made of the original Euclidean space $\mathbb{R}^d$ supplemented by two Grassmann dimensions ($\theta,\bartheta$), which we collectively denote by $\underline{\theta}$. Here, $\tilde h=ih-i{g_0\over2}\{\bar c,c\}$ and the $x$-dependence only appears through the fields $U$, $c$, $\bar c$ and $h$. It is straightforward to show that 
\begin{equation}
  \label{eq_susy2}
S_{\rm \!GF}[A,{\cal V}] =   \int_{x,\underline{\theta}}\tr\left\{\left({\cal D}_\mu{\cal V}\right)^{\!\dagger}\!\left({\cal D}_\mu{\cal V}\right)\right\},
\end{equation}
where we introduced the covariant derivative ${\cal D}_\mu{\cal V}\equiv\partial_\mu{\cal V}+ig_0{\cal V}\!A_\mu$ and $\int_{\underline{\theta}}\equiv  \int d\theta d\bartheta\,g^{1/2}(\underline{\theta})$, with $g^{1/2}(\underline{\theta})=\left(\beta_0\thetab\theta-1\right)$,
can be seen as the invariant measure associated with curved Grassman dimensions \cite{Tissier_08}. \Eqn{eq_susy2} is the action of a supersymmetric gauged NL$\sigma$ model on a curved superspace. It is invariant under the isometries of the curved superspace, the super gauge transformation $A_\mu\to A_\mu^{\cal U}={\cal U}A_\mu{\cal U}^{-1}+{i\over g_0}{\cal U}\partial_\mu{\cal U}^{-1}$ and ${\cal V}\to{\cal V}{\cal U}^{-1}$, where ${\cal U}\equiv{\cal U}(x,\underline{\theta})$ is a local element of $SU\!(N)$ on the superspace, as well as under the right $SU\!(N)$ transformation ${\cal V}\to{\cal U}_R{\cal V}$, where the matrix ${\cal U}_R\equiv{\cal U}_R(\underline{\theta})$ can be local in Grassmann variables.

A non-trivial issue in \Eqn{eq_av_A} is the presence of  the denominator of the average \eqn{eq:fact1} over Gribov copies, which is a highly non-linear, non-local functional of the gauge field $A$. Similar two-step averagings are common in the theory of disordered system and are efficiently dealt with by means of the replica trick \cite{young}. In its simpler version, the latter amounts to writing, formally,
\begin{equation}
  \label{eq_replica1}
  \frac{1}{\int \mathcal D {\cal V}
    \, e^{-S_{\rm \!GF}[A,{\cal V}]}}=\lim_{n\to0}\int\prod_{k=1}^{n-1}\left( \mathcal D {\cal V}_k
    \, e^{-S_{\rm \!GF}[A,{\cal V}_k]}\right)\,.
\end{equation}
Introducing a $n$-th replica from the numerator in \eqn{eq_av_A}, the final average over gauge fields then reads
\begin{equation}
  \label{eq_average2bis}
  \overline{\langle{\cal O}[A]\rangle}=\lim_{n\to 0}\frac{\int\mathcal D A\left(\prod_{k=1}^n \mathcal D {\cal V}_k\right)\,{\cal O}[A^{U_n}]\, e^{-S[A,\{{\cal V}\}]}}{\int\mathcal D A\left(\prod_{k=1}^n \mathcal D {\cal V}_k\right)\, e^{-S[A,\{{\cal V}\}]}}\,,
\end{equation}
where $S[A,\{{\cal V}\}]=S_{\rm YM}[A]+\sum_{k=1}^{n}S_{\rm \!GF}[A,{\cal V}_k]$.
For perturbative calculations, one needs to factor out the volume of the gauge group. This can be done by exploiting the gauge invariance of the integration measure ${\cal D}A$ and of the YM action and using appropriate changes of variables to extract, say, a factor $\int {\cal D} U_n$. Renaming $(c_n,\cb_n,h_n)\to (c,\cb,h)$, one finally gets
\begin{equation}
  \label{eq_average2ter}
  \overline{\langle{\cal O}[A]\rangle}=\lim_{n\to 0}\frac{\int\mathcal D (A,c,\bar c,h,\{{\cal V}\})\,{\cal O}[A]\, e^{-S[A,c,\bar c,h,\{{\cal V}\}]}}{\int\mathcal D (A,c,\bar c,h,\{{\cal V}\})\, e^{-S[A,c,\bar c,h,\{{\cal V}\}]}}\,,
\end{equation}
with $\mathcal D (A,c,\bar c,h,\{{\cal V}\})\equiv\mathcal D (A,c,\bar c,h)\left(\prod_{k=1}^{n-1} \mathcal D {\cal V}_k\right)$ and
\beq
\label{eq:action}
  S[A,c,\bar c,h,\{{\cal V}\}]=\int_x\tr \left\{{1\over2}F^2+{\beta_0}A^2+2\partial_\mu\bar cD_\mu c+2ih\partial_\mu A_\mu\right\}+{1\over g_0^2}\sum_{k=1}^{n-1} \int_{x,\underline{\theta}}\tr \left\{\!\left({\cal D}_\mu{\cal V}_k\right)^{\!\dagger}\!\left({\cal D}_\mu{\cal V}_k\right)\!\right\}\!.
\eeq
This describes a collection of $n-1$ gauged supersymmetric NL$\sigma$ models coupled to a gauge-fixed YM field with gauge fixing $S_{\rm GF}[A,\mathds{1},c,\cb,h]$. Notice that, as is clear from Eqs. \eqn{eq:gf} and \eqn{eq_susy2}, each replica produces a term $\beta_0 A^2$, thus giving rise to a bare mass term $n\beta_0A^2$. Thanks to its large number of symmetries, the theory \eqn{eq:action} has been shown to be perturbatively renrormalizable in $d=4$, with only two renormalization factors, in \cite{Serreau:2012cg}.

\section{Perturbative equivalence with the Landau limit of the Curci-Ferrari model}

Perturbation theory is most conveniently formulated in the supersymmetric formalism, which makes transparent the (dramatic) consequences of the supersymmetries for loop diagrams. Pa\-ra\-me\-tri\-zing the constrained superfields ${\cal V}_k$ in terms of unconstrained ones, e.g., $\mathcal V_k=\exp\left\{{ig_0\Lambda_k}\right\}$, it is straightforward to obtain the various propagators of the theory \eqn{eq:action}, written below in Euclidean momentum space. The gluon propagator reads:
 \begin{equation}
  \label{eq_propagAA}
   \left[ A^a_\mu(p)\,A^b_\nu(-p)\right]=\frac{\delta^{ab}}{p^2+n\beta_0}\left(\delta_{\mu\nu}-\frac{p_\mu
      p_\nu}{p^2}\right),
\end{equation}
where the square brackets represent averaging with the action (\ref{eq:action}) with $n\neq0$.
It is transverse in momentum, as a result of Landau gauge, and massive, with bare square mass $n\beta_0$, as a result of our particular gauge fixing procedure. The ghost propagator assumes the usual form:
\beq
   \label{eq_propagcc}
\left[ c^a(p)\,\cb^b(-p)\right]={\delta^{ab}}/{p^2}\,.
\eeq
Finally, the superfield propagator is given by
\begin{equation}
  \label{eq_propagLL}
  \left[\Lambda^a_k(p,\underline{\theta})\,\Lambda^b_l(-p,\underline{\theta}')\right]=\delta^{ab}\,\delta_{kl}\,\delta(\underline\theta,\underline\theta')/{p^2}\,,
\end{equation}
where $\delta(\underline\theta,\underline\theta')=g^{-1/2}(\underline{\theta})\,(\thetab-\thetab')(\theta-\theta')$ is the Dirac function on the curved Grassman space.

The vertices of the action \eqn{eq:action} which do not involve the superfields $\Lambda_k$ are the same as for the usual FP Landau gauge. The NL$\sigma$ model sector of the theory gives vertices with an arbitrary number of $\Lambda_k$ legs and either zero or one gluon leg. Clearly, the latter are local in Grassmann variables. Using \eqn{eq_propagLL}, we conclude that any closed loop of $\Lambda_k$ superfields with $p$ vertices insertions involves the product $
  \delta(\underline\theta_1,\underline\theta_2)\cdots\delta(\underline\theta_p,\underline\theta_1)\propto \delta(\underline\theta_1,\underline\theta_1)=0$.
Hence, the NL$\sigma$ model sector of the theory is tree-level exact. This is not surprising since, in the above construction, the role of these superfields is in fact to reconstruct the weighted sum \eqn{eq_average_G} on Gribov copies, that is to project on the extrema of the action \eqn{eq_z}.

This observation has two important consequences. First, the only perturbative source of dependence in the number $n$ of replicas is the bare gluon mass in \eqn{eq_propagAA}. Second, correlators or vertex functions involving only the fields $A$, $c$ and $\cb$ do not receive any loop contribution from superfields. They are thus obtained from the very same diagrams as in the FP Landau gauge, with usual YM vertices and with propagators given by \eqn{eq_propagAA} and \eqn{eq_propagcc}, that is from the effective action
\beq
\label{eq:actioneff}
  S[A,c,\bar c,h,\{{\cal V}\}]\to S_{\rm eff}[A,c,\bar c,h]=\int_x\tr \left\{{1\over2}F^2+{n\beta_0}A^2+2\partial_\mu\bar cD_\mu c+2ih\partial_\mu A_\mu\right\}.
\eeq
This simple massive extension of the FP action is a particular case of the CF model \cite{Curci76}. 

A key point here concerns the issue of the limit $n\to0$ versus renormalization. It is clear from the above considerations that taking first the zero replica limit leads to the standard FP Landau gauge. This, however, is not satisfactory since one expects gauge-dependent quantities to depend on the gauge-fixing parameter $\beta_0$. A similar situation arises if one introduces a renormalized gauge-fixing parameter, e.g., as $\beta_0=Z_\beta \beta$. An alternative renormalization scheme is to redefine the square mass as $n\beta_0=Z_{m^2} m^2$. This absorbs the remaining $n$-dependence and gives a non-trivial $n\to0$ limit.

\section{One-loop results}

\begin{figure}[h]
  \centering
  \includegraphics[width=.4\linewidth]{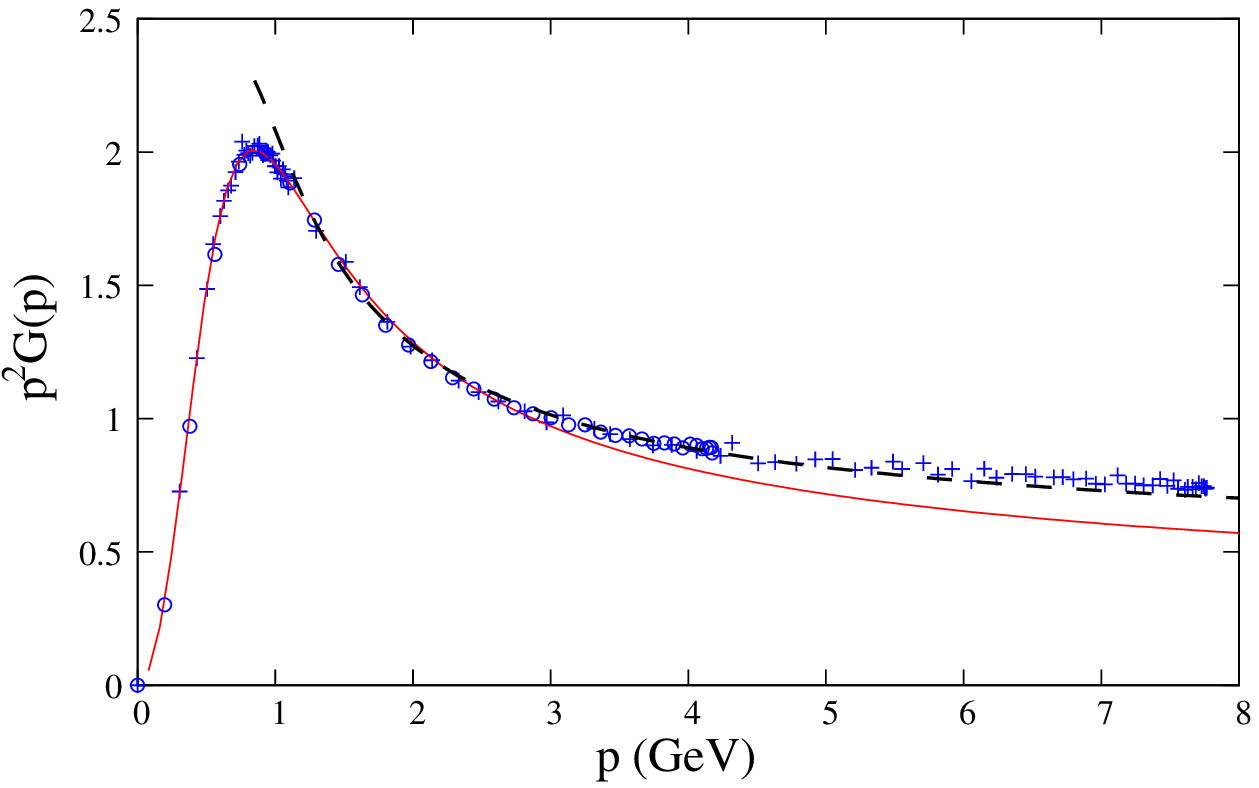}\qquad
   \includegraphics[width=.4\linewidth]{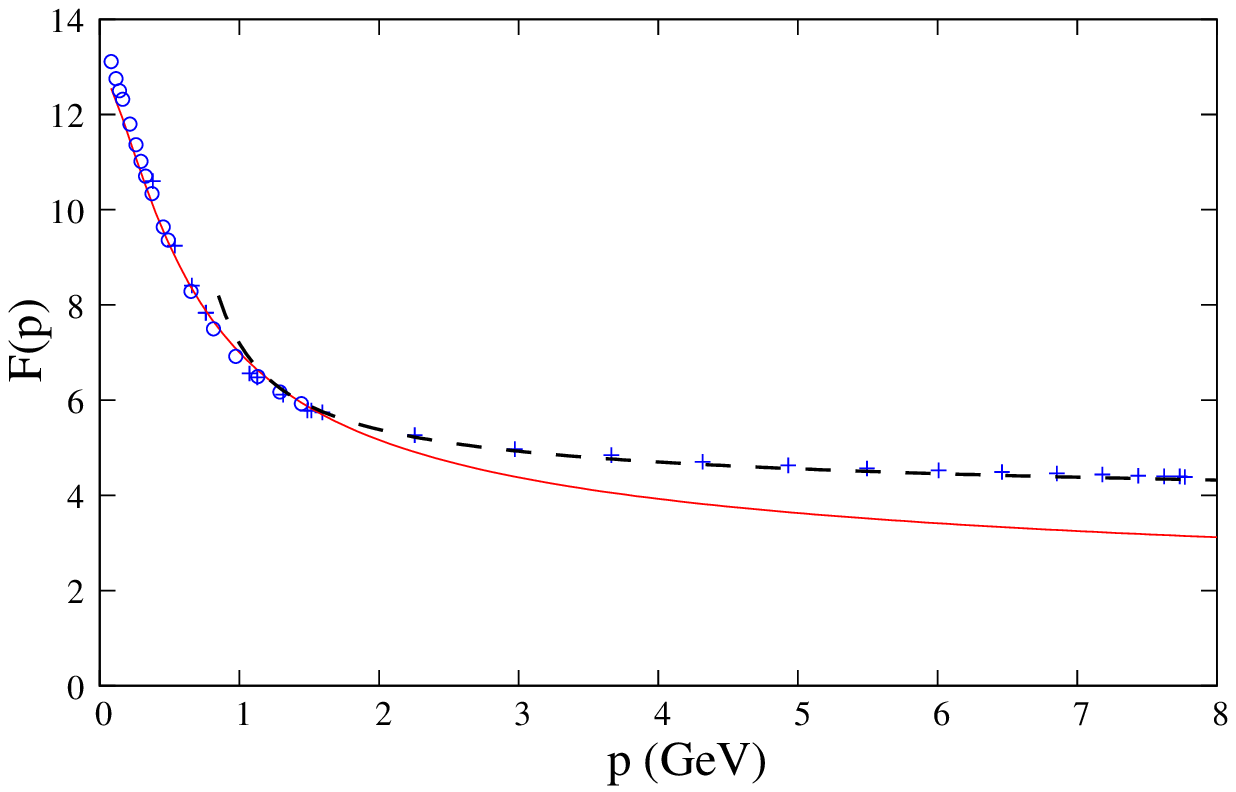}
 \caption{The gluon and ghost dressing functions, $p^2G(p)$ and $F(p)=p^2D(p)$, for the $SU\!(3)$ theory in $d=4$, obtained in [11]. Lattice results are shown as blue dots. The lines show the one-loop results obtained from the action (3.4), without (red) and with (black dashes) renormalization group improvement in the UV. The employed renormalization conditions are $G^{-1}(0)=m^2$, $G^{-1}(\mu)=m^2+\mu^2$, $D^{-1}(\mu)=\mu^2$ and a standard Taylor scheme for the coupling. The parameters are $m=0.54$ GeV and $g=4.9$ at $\mu=1$ GeV.}
  \label{fig_prop}
\end{figure}

The ghost and gluon propagators have been computed at one-loop in the theory \eqn{eq:actioneff} in \cite{Tissier_10}. The results are in remarkably good quantitative agreement with lattice data for $SU\!(2)$ and $SU\!(3)$ in $d=4$ and give a fairly good qualitative description in $d=3$. Fig. \ref{fig_prop} reproduces the gluon and ghost propagators $G_{\mu\nu}^{ab}(p)=\delta^{ab}(\delta_{\mu\nu}-p_\mu p_\nu/p^2)G(p)$, and $D^{ab}(p)=\delta^{ab} D(p)$ computed in  \cite{Tissier_10}.

Another interesting observation is that the theory \eqn{eq:actioneff} admits infrared safe renormalization schemes, with no Landau pole, as discussed in \cite{Tissier_10,Serreau:2012cg}.\footnote{The IR flow of this theory has also been studied in \cite{Weber:2011nw}.} Defining the renormalized fields and constants $A= \sqrt{Z_A} A_r$, $c= \sqrt{Z_c} c_r$, $\bar c= \sqrt{Z_c} \bar c_r$, $g_0= Z_g g$ and $n\beta_0= Z_{m^2} m^2$, the following renormalization conditions
\beq
\label{eq:scheme}
 G^{-1}(\mu)=m^2+\mu^2,\qquad D^{-1}(\mu)=\mu^2,\qquad Z_A\,Z_c\,Z_{m^2}=1,
\eeq
together with the Taylor condition for the coupling, $Z_gZ_c\sqrt{Z_A}=1$, lead to the renormalization group (RG) flow depicted in Fig. \ref{fig_flow}. 
There is an ultraviolet (UV) attractive fixed point at $m=0$ and $g=0$: both the bare coupling $g_0$ and the bare mass $n\beta_0$ vanish in the process of removing the UV regulator. We note that this is compatible with taking the limit $n\to0$ at fixed gauge-fixing parameter $\beta_0$. 
Most remarkably, RG trajectories fall in two distinct classes, depending on the initial conditions in the UV: those with or without a Landau pole in the IR. It is worth mentioning that the best values of $m$ and $g$ for describing lattice data in the scheme \eqn{eq:scheme} belong to the second class~\cite{Tissier_10}.

\section{Issues and perspectives}

\begin{figure}[t]
  \centering
  \includegraphics[width=.37\linewidth]{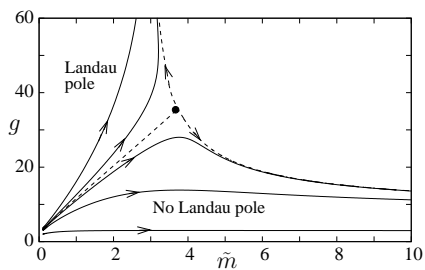}
  \caption{One-loop RG flow in the plane ($\tilde m=m/\mu,g$). The arrows indicate the flow towards the infrared.}
  \label{fig_flow}
\end{figure}

The gauge-fixing proposed in \cite{Serreau:2012cg} provides what appears to be an essential feature of the gluon propagator, an effective mass term, already at tree level. This raises the following issues. In principle, different values of the gauge-fixing parameter $\beta_0$, corresponding to different RG trajectories in Fig. \ref{fig_flow}, could give (vastly) different results for the YM correlators. Instead, lattice results in the minimal Landau gauge show at best a mild sensitivity with the selected Gribov copy \cite{Sternbeck:2005tk}. However, it is conceivable, as mentioned previously, that the minima of ${\cal F}[A^U]$ be nearly degenerate and well separated from the first saddles. This is supported by numerical investigations on small lattices, where all copies can actually be found \cite{Hughes:2012hg}. If so, there exists some range of values of $\beta_0$ which corresponds to giving essentially an equal weight to copies in the first Gribov region --probed by lattice simulations-- and to suppressing those outside the first region. It is presently not known how lattice results are affected if one selects a copy outside the first Gribov region. This issue requires detailed studies of Gribov copies, including sadles, in the spirit of \cite{Hughes:2012hg}.

Another important question concerns the limit $n\to0$ versus renormalization which, as already mentioned, do not commute. A similar issue arises in the theory of disordered systems, where one is concerned with the thermodynamic limit instead of renormalization. In this context, it is understood that the limit $n\to0$ should always be taken last \cite{young}. In fact, replicas can be viewed as non-trivial external fields which allow one to probe the complicated landscape of the potential and which should be removed only at the end of the calculations in order to probe non trivial physics, such as spontaneous symmetry breaking. We believe this picture can be fruitful in the YM context too. Sending $n\to0$ naively leads to the standard FP Landau gauge with nilpotent BRST symmetry. Taking, instead, the limit $n\to0$ after having properly removed the UV regulator, one captures non-trivial IR physics where, however, the nilpotent BRST symmetry is broken. This is very transparent after the superfields have been integrated out in \eqn{eq:actioneff}. The mass term $n\beta_0$ can be seen as an external source which explicitly breaks the BRST symmetry. The fact that the symmetry is not recovered after this source is eventually removed signals that it is spontaneously broken by IR fluctuations.\footnote{We mention that a similar construction can be made for QED. In this case, however, the mass term does receive nontrivial renormalization and thus vanishes in the limit $n\to0$.}

To conclude, we believe the proposal of \cite{Serreau:2012cg} opens a possibility to access non-trivial IR physics by perturbative methods. Ongoing research in this context includes the calculation of higher-order corrections and higher-order vertices in the theory \eqn{eq:actioneff} and the inclusion of quarks \cite{prep2}, or the calculation of YM correlators and thermodynamics at finite temperature \cite{prep1}. It may be of interest to investigate the non-perturbative aspects of the theory \eqn{eq:action} either with continuum methods or with lattice simulations, in the spirit of \cite{Henty:1996kv}, see also \cite{Maashere}. Other interesting questions concern the the generalization of the approach of \cite{Serreau:2012cg} to other gauges \cite{prep0}, or the relation with other proposals such as, e.g., \cite{Maas:2009se}. Finally, a major open question is that of unitarity \cite{Kondo:2012ri}.

\section*{Acknowledgements}

I would like to thank the organizers for setting up this enjoyable and stimulating conference. I acknowledge interesting discussions with A. Accardi, C. Fischer, J. Greensite, K.-I. Kondo, A. Maas, J. M. Pawlowski and D. Zwanziger. I am also grateful to U. Reinosa, M. Tissier and N. Wschebor for the enriching collaboration on these topics.

\end{document}